\documentclass{JINST}
\usepackage{epsfig}
\usepackage{subfigure}
\usepackage{here}
\usepackage{graphicx}
\usepackage{mathtools}

\makeatletter
\renewcommand{\@thesubfigure}{\Large(\alph{subfigure})}
\renewcommand{\p@subfigure}{figure\space}
\renewcommand{\p@figure}{figure\space}
\makeatother %

\title{Measurement of the absolute Quantum Efficiency of Hamamatsu model R11410-10 photomultiplier tubes at low temperatures down to liquid xenon boiling point.}

\author{A. Lyashenko$^a$\thanks{Corresponding author.}~,
T. Nguyen$^a$, A. Snyder$^b$, H. Wang$^a$ and K. Arisaka$^a$\\
\llap{$^a$}University of California Los Angeles,\\
  Department of Physics and Astronomy, 475 Portola Plaza, Los Angeles CA 90095 USA\\
\llap{$^b$}University of Illinois at Urbana-Champaign,\\
    1110 West Green Street, Urbana IL 61801 USA\\
E-mail: \email{lyashen@physics.ucla.edu}}

\abstract{We report on the measurements of the absolute Quantum Efficiency(QE) for Hamamatsu model R11410-10 PMTs specially designed for the use in low background liquid xenon detectors. QE was measured for five PMTs in a spectral range between 154.5 nm to 400 nm at low temperatures down to -110$^0$C. It was shown that during the PMT cooldown from room  temperature to -110 $^0$C (a typical PMT operation temperature in liquid xenon detectors), the absolute QE increases by a factor of 1.1 - 1.15 at 175 nm. The QE growth rate with respect to temperature is wavelength dependent peaking at about 165 nm corresponding to the fastest growth of about -0.07 \%QE/$^{0}C$ and at about 200 nm corresponding to slowest growth of below -0.01 \%QE/$^{0}C$. A dedicated setup and methods for PMT Quantum Efficiency measurement at low temperatures are described in details. }


\begin{document}

\section{Introduction}\label{sec:intro}
The R11410 type Photo-Multiplier Tube (PMTs) was designed by Hamamatsu and developed in a close collaboration with UCLA \cite{bib1},\cite{ham_talk} for reading out scintillation signals in extremely low radioactive background liquid xenon experiments \cite{bib4}, \cite{lux}, \cite{xenon1t}. To reduce both electron and nuclear recoil backgrounds in the active detection volume, the R11410 PMTs were made of extremely low radioactivity materials. The overall PMT weight was also reduced to about 200 g. The radioactivities quoted by the manufacturer are: 3.3 mBq/piece for $^{238}$U, 2.3 mBq/piece for $^{232}$Th, 5.7 mBq/piece for $^{40}$K and 9.1 mBq/piece for $^{60}$Co. Even lower intrinsic radioactivities have been recently reported for a newer variant of the R11410 \cite{bib2}. Higher sensitivity to light signals due to improved photocathode material and photo-electron collection as well as low dark count rate will ensure a lower $<$10 keVnr nuclear recoil equivalent energy detection threshold for primary scintillation in liquid xenon detectors. Due to a larger active area the number of PMT units used in the detector and therefore its overall cost could be significantly reduced without sacrificing considerably the spatial resolution of the detector.

Understanding the PMT response to various light signals produced in a liquid xenon detector is a critical part of the experiment as it directly affects experimental results. Therefore, measuring such PMT characteristics as Quantum Efficiency (QE), gain, dark count rate, linearity of the response, photocathode uniformity and afterpulsing is very important especially at liquid xenon temperature as some of those characteristics could be temperature dependent. Therefore it is also important to measure the above mentioned parameters at liquid xenon temperature. Gain, dark count rate, linearity of the response, photocathode uniformity and afterpulsing for the R11410-10 PMT were previously measured \cite{bib1}. Performance of the R11410 PMT was studied in a liquid xenon environment demonstrating stability of gain in the vicinity of a strong electric field \cite{bib3}. However, measurements of Quantum Efficiency of the R11410-10 PMT at liquid xenon temperature, have not been reported yet. The relative QE in liquid xenon was previously measured for the Hamamatsu model R8520 PMTs \cite{aprile:12} and the ETL D730/9829Q PMTs \cite{araujo:03} used in XENON100 and ZEPLIN III dark matter experiments correspondingly, showing some 5-20\% increase in QE at -100 $^0$C compared to that at room temperature. Some earlier studies performed with various PMTs \cite{araujo:97}, \cite{araujo:98} also indicated at the temperature dependance of the PMT response to light signals. All these PMTs have bi-alkali photocathodes, therefore, one could expect some increase in QE for R11410 PMTs operated in similar temperature conditions. R8520 and R11410 PMTs are both equipped with high sensitivity bi-alkali photocathodes reaching QEs in excess of 30\% near the peaks values at 175 nm and at 340 nm due to an improved photocathode production process employed by Hamamtsu Photonics. Similar photocathode characteristics at 340 nm were also achieved earlier in the laboratory \cite{lyas:09:1}, \cite{lyas:09:2}.

In this article we report on our measurements of the absolute QE for Hamamatsu model R11410-10 PMT at low temperatures down to liquid xenon temperature (-110$^0$C). We describe in details the experimental methods and discuss the results.

\section{Experimental}\label{sec:exper}
\begin{figure}[tbp] % figures (and tables) should go top or bottom of
                    % the page where they are first cited or in
                    % subsequent pages
\begin{center}%
\subfiguretopcaptrue
\subfigure[][] % caption for subfigure a
{
\label{subfig:setup}
\includegraphics[width=9cm]{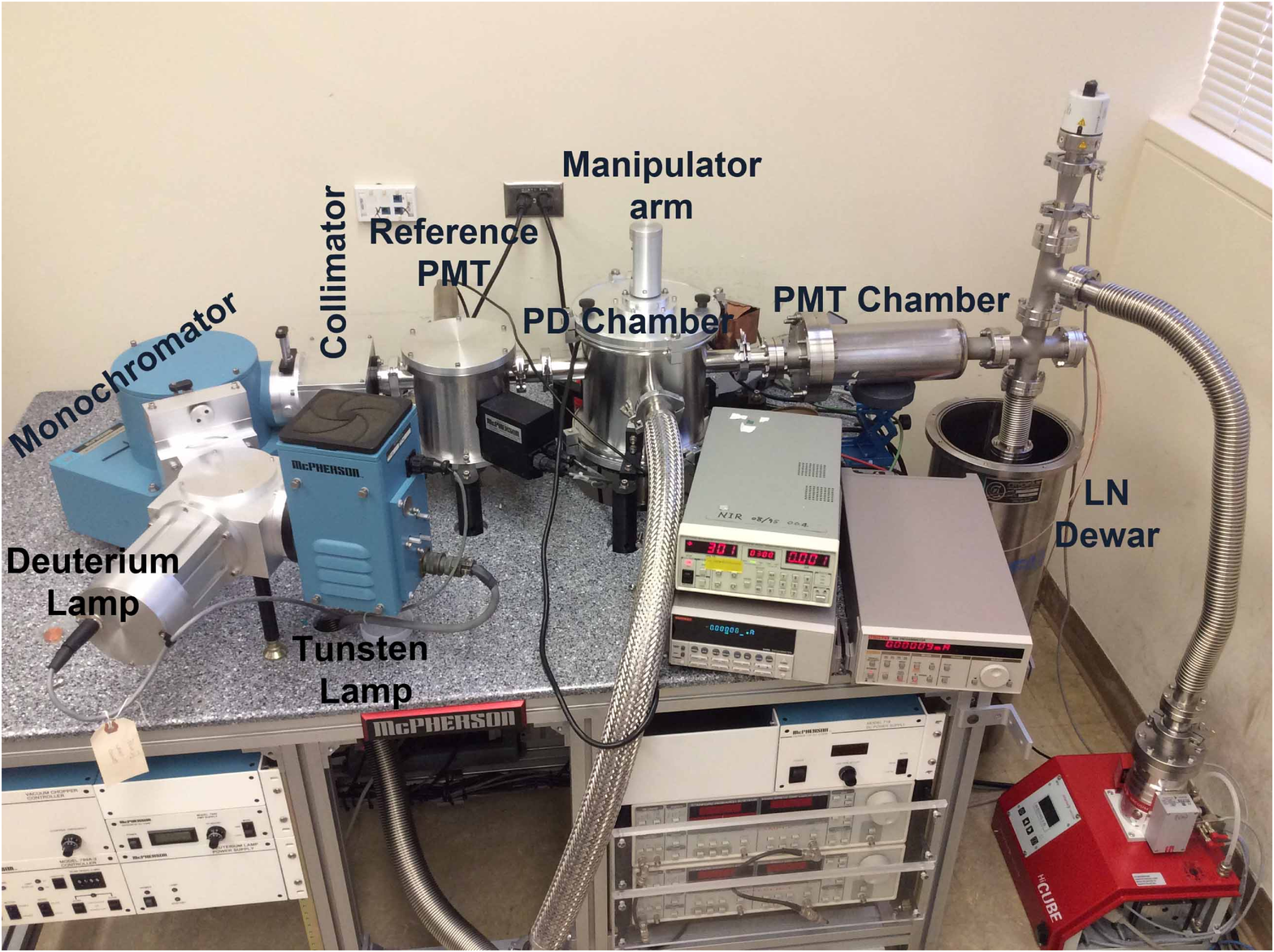}
}
\subfigure[][]
{
\label{subfig:scheme}
\includegraphics[width=9cm]{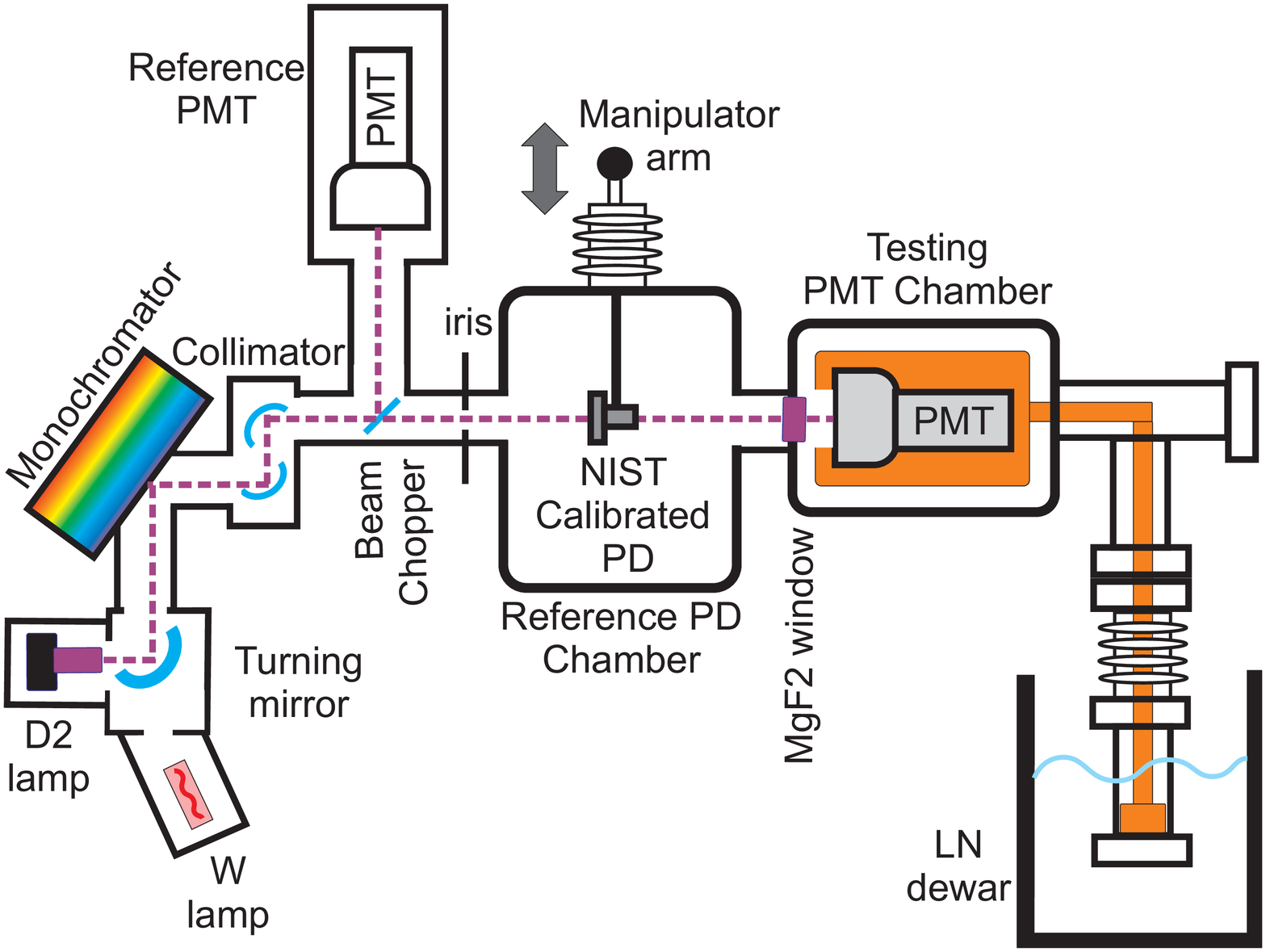}
}
\caption{A Photograph of the experimental setup for measurement of the absolute QE of R11410 PMT at cryogenic temperature, b) schematic view of the setup.}
\label{fig:exp_set}
\end{center}%
\end{figure}

The experimental method used for the absolute QE measurements was based on the comparison of photocurrents of the test PMT with that of a reference Photo-Diode \footnote{Opto Diode Corp. model AXUV-100G} calibrated by the National Institute of Standards and Technology (NIST). The experimental setup for measurement the absolute QE shown in \ref{fig:exp_set} was comprised of the test PMT vacuum chamber(hereinafter PMT Chamber) attached to a vacuum spectrometer\footnote{McPherson model VUVAS 2000}. Vacuum volumes of the spectrometer and the PMT chamber were separated using a flange with $MgF_2$ window.

The reference Photo-Diode (PD) was fixed on a linear motion manipulator arm in the sample chamber (hereinafter PD Chamber) of the spectrometer. Each vacuum volume (PMT Chamber and spectrometer) was evacuated with a separate turbo-molecular pump. The PMT Chamber was evacuated to high vacuum of 3$\cdot10^{-7}$ mbar to avoid water condensate on the PMT window during cooldown; the spectrometer was evacuated to 3$\cdot10^{-4}$ mbar to ensure good light transmittance in VUV spectral range along the light path (see \ref{subfig:scheme}). We utilized a deuterium lamp as a light source for the QE measurements in a range from 154.5 nm to 400 nm; the Tungsten lamp was used for measurements in a range from 300 nm to 600 nm. One could switch between the light sources using a rotating mirror as shown in \ref{subfig:scheme}. Light from either light source was fed into the entrance slit of the spectrometer. Then collimated monochtromatic light beam was split equally in two beams with a beam chopper. The deflected part of the beam was fed onto a reference PMT while the rest of the beam was focused onto the NIST calibrated PD. The diameter of the illuminated spot on the PD with an active area of 10x10 $mm^{2}$ was set to 7 mm using the iris at the output of the spectrometer to ensure full light collection by the PD. The PD could be moved out of the light path with the manipulator arm. In this case the light flux delivered to the test PMT is equal to that delivered to the reference PD times the transmission coefficient of the $MgF_{2}$ window as shown below.

\begin{figure}[tbp] % figures (and tables) should go top or bottom of
                    % the page where they are first cited or in
                    % subsequent pages
\begin{center}%
\includegraphics[width=4cm]{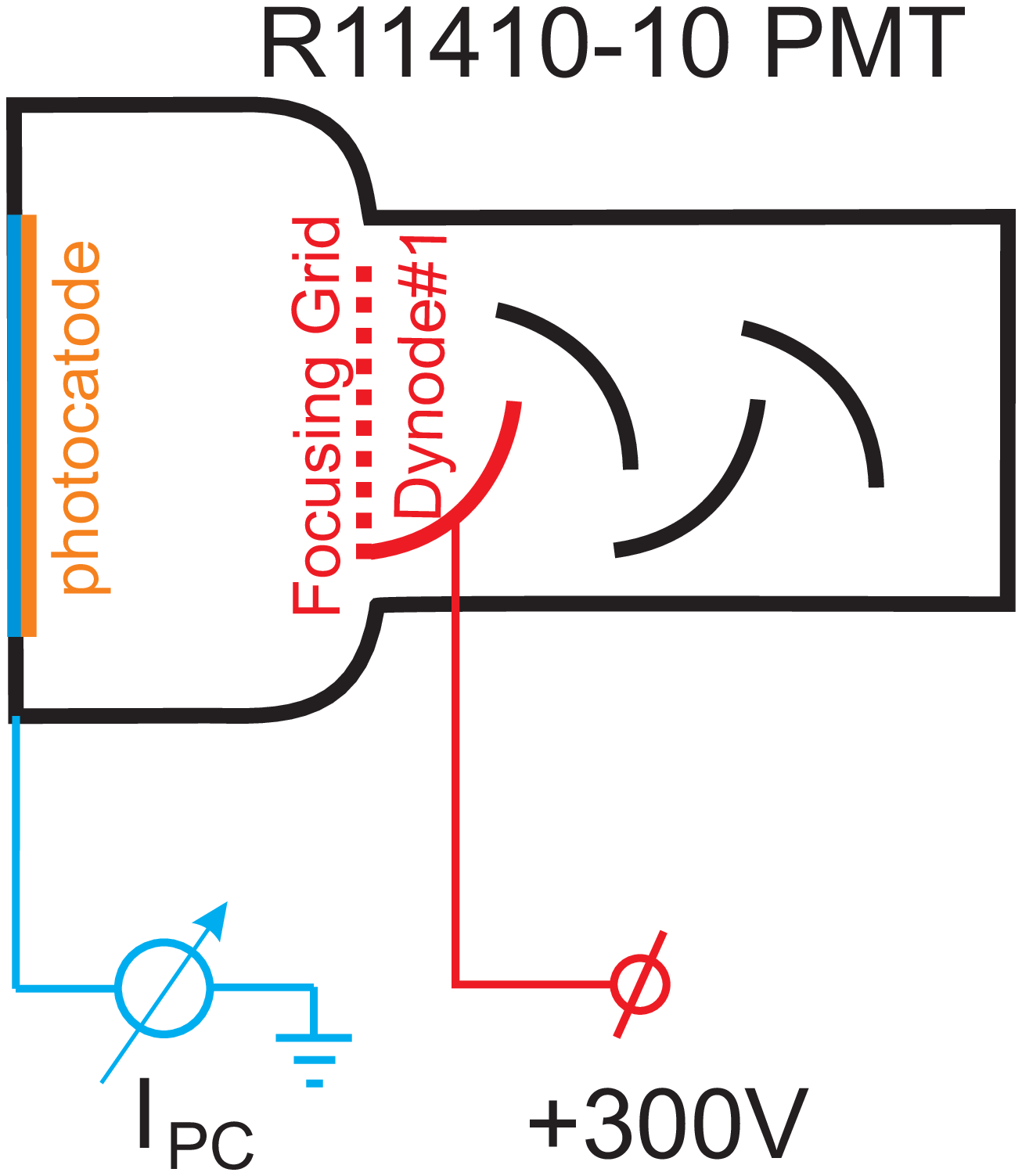}
\caption{The PMT electrical connection scheme.}
\label{fig:pmt_connection}
\end{center}%
\end{figure}

Currents from either reference \footnote{Hamamatsu model R6835} or test PMT was read out at the PMT cathode using picoammeters\footnote{Keithley model 486 and Keithley model 6485 correspondingly}.  Both reference and test PMTs were operating in a DC current mode by applying a positive bias of 300 V at the first dynode to extract photoelectrons as shown in \ref{fig:pmt_connection}. For the R11410-10 PMT, the first dynode is interconnected with the focusing grid inside the PMT inclosure (\ref{fig:pmt_connection}). The PD current was read out with a picoammeter\footnote{Keithley model 6485} as well. The picoammeters were connected to a PC via a GPIB interface. To account for possible fluctuations of the deuterium lamp light intensity over time, the current readings were normalized to the current read on the reference PMT (see \ref{fig:exp_set}). The input window of the reference PMT was coated with Tetraphenyl butadiene (TPB) wavelength shifter for the UV operation. Measurements were fully automated using LabView software. Currents were recorded whilst varying the wavelength on the spectrometer in a range from 154.5 nm to 400 nm for the deuterium lamp and from 300 nm to 600 nm for the tungsten lamp. To improve systematics each current reading was averaged over 50 measurements. An integration time for each current reading was 100 msec. That corresponds to the measurement time of 5 sec spent at each wavelength. As shown below, a positive bias of 300V ensures nominal photoelectron collection efficiency. The latter was defined as a ratio of the photocurrent measured at the first dynode of the PMT to the photocurrent measured at the photocathode of the PMT while applying the same voltage difference between the PMT cathode and the first dynode. The photocurrent at the first dynode was measured using an inverted electrical connection scheme compared to \ref{fig:pmt_connection} by applying a negative bias to the photocathode and reading out the current directly at the first dynode. The photocurrents were normalized by the current of the reference PMT to account for possible fluctuations in the light. The photoelectron collection efficiency was calculated using Eq. \ref{eq:pe_collection}:

\begin{equation}
\label{eq:pe_collection}
\varepsilon(\Delta V_{CDy1})=\frac{I_{Dy1}(\Delta V_{CDy1})-I_{Dy1}^{dark}}{I_{PC}(\Delta V_{CDy1})-I_{PC}^{dark}} \cdot \frac{I_{ref}^{'}-I_{ref}^{dark}}{I_{ref}-I_{ref}^{dark}}
\end{equation}

\noindent where

$\Delta V_CDy1$ - absolute voltage difference between the photocathode and the first dynode

$\varepsilon(\Delta V_{CDy1})$ - photoelectron collection efficiency

$I_{Dy1}(\Delta V_{CDy1})$ - photocurrent measured on the first dynode of the test PMT

$I_{Dy1}^{dark}$ - dark-current measured on the first dynode of the test PMT with the spectrometer lamp switched off

$I_{PC}(\Delta V_{CDy1})$ - photocurrent measured on the photocathode of the test PMT

$I_{PC}^{dark}$ - dark-current measured on the photocathode of the test PMT with the spectrometer lamp switched off

$I_{ref}^{'}(\lambda)$ - current on the reference PMT associated with $I_{PC}(\Delta V_{CDy1})$ measurements

$I_{ref}(\lambda)$ - current on the reference PMT associated with $I_{Dy1}(\Delta V_{CDy1})$ measurements

$I_{ref}^{dark}$  - dark-current on the reference PMT with the spectrometer lamp switched off

The photoelectron collection efficiency calculated using Eq. \ref{eq:pe_collection} should be corrected by the optical transparency of the PMT focussing grid (\ref{fig:pmt_connection}). Recent electrostatic simulations of the R11410 PMT suggest that the optical transparency of the focusing grid is about 95\% \cite{mayani:14}.

The absolute QE at a given temperature was measured in two steps. On the first step the PD current together with the reference PMT currents were read out while scanning through a spectral range of interest. On the second step the current at the test PMT and that at the reference PMT were recorded in the same spectral range. Each step took about 2 minutes totaling at about 4 minutes for the absolute QE measurement at a given temperature in a given spectral range. During each 4 minute scan the PMT temperature was varying by less then 2 $^0$C as the PMT cooling process was rather slow. Switching between various light sources was very time consuming procedure as it requires replacing the diffractive gratings in the spectrometer. Because of that, the absolute QE at low temperatures (below room temperature) was measured only with the deuterium lamp in a range from 154.5 nm to 400 nm. At room temperature the absolute QE was measured with two light sources deuterium lamp and tungsten lamp covering the spectral range from 154.5 nm to 600 nm.

\begin{figure}[tbp] % figures (and tables) should go top or bottom of
                    % the page where they are first cited or in
                    % subsequent pages
\begin{center}%
\includegraphics[width=9cm]{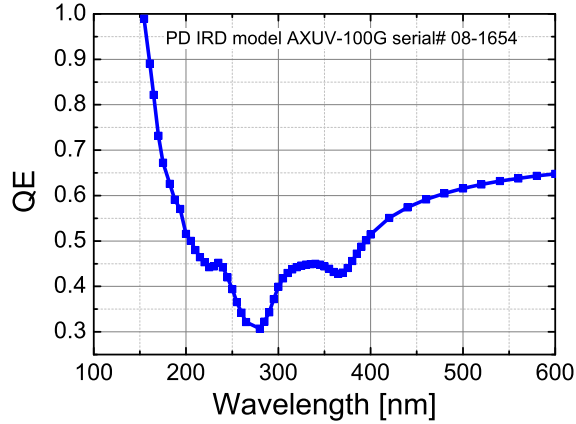}
\caption{QE (the average number of photoelectrons per incident photon) of the reference Photodiode as provided in the NIST calibration report.}
\label{fig:qe_pd}
\end{center}%
\end{figure}

The absolute QE of the PMT was calculated as follows:
\begin{equation}
\label{eq:qe}
QE(\lambda)=\frac{I_{PMT}(\lambda)-I_{PMT}^{dark}}{I_{PD}(\lambda)-I_{PD}^{dark}} \cdot \frac{I_{ref}^{'}(\lambda)-I_{ref}^{dark}}{I_{ref}(\lambda)-I_{ref}^{dark}}\cdot \frac{QE_{PD}(\lambda)}{T_{MgF_2}(\lambda)}
\end{equation}

\noindent where

$\lambda$ - wavelength in nm

$QE(\lambda)$ - test PMT quantum efficiency

$I_{PD}(\lambda)$ - photocurrent measured on the PD

$I_{PD}^{dark}$ - dark-current measured on the photocathode with the spectrometer lamp switched off

$I_{PMT}(\lambda)$ - current measured on the test PMT

$I_{PMT}^{dark}$ - dark-current measured on the test PMT with the spectrometer lamp switched off

$I_{ref}^{'}(\lambda)$ - current on the reference PMT from the PD measurement

$I_{ref}(\lambda)$ - current on the reference PMT from the test PMT measurement

$I_{ref}^{dark}$  - dark-current on the reference PMT with the spectrometer lamp switched off

$QE_{PD}(\lambda)$ - mean PD quantum efficiency as supplied by NIST (see \ref{fig:qe_pd})

$T_{MgF_2}(\lambda)$ - $Mg_{2}F$ window light transmission (see \ref{fig:mgf2})

We should make a remark here that the absolute QE at room temperature was measured without the $MgF_{2}$ window by setting $T_{MgF_2}(\lambda)$=1 in Eq. \ref{eq:qe}. The window was only used for the QE measurements at low temperatures to improve vacuum conditions near the PMT by separating the PMT Chamber from the PD Chamber (see \ref{fig:exp_set}). As mentioned above a bad vacuum near the PMT would have resulted in a deposition of a water condensate onto the PMT window making it opaque to Ultra-Violet light.

To measure $T_{MgF_2}(\lambda)$ the current was recorded at the test PMT either with and without the $MgF_{2}$ window. Transmission $T_{MgF_2}(\lambda)$  was calculated as a ratio of the test PMT current measured with the $MgF_{2}$ $I_{PMTMgF_2}(\lambda)$ to that measured without the window $I_{PMT}(\lambda)$  normalized by the current recorded at the reference PMT $I_{ref}(\lambda)$ as shown in Eq. \ref{eq:trans}.

\begin{equation}
\label{eq:trans}
T_{MgF_2}(\lambda)=\frac{I_{PMTMgF2}(\lambda)-I_{PMT}^{dark}}{I_{PMT}(\lambda)-I_{PMT}^{dark}} \cdot \frac{I_{ref}^{'}(\lambda)-I_{ref}^{dark}}{I_{ref}(\lambda)-I_{ref}^{dark}}
\end{equation}
\noindent where

$T_{MgF_2}(\lambda)$ - $Mg_{2}F$ window light transmission

$I_{PMTMgF2}(\lambda)$ - current measured on the test PMT with $MgF_{2}$ window

$I_{PMT}(\lambda)$ - current measured on the test PMT without $MgF_{2}$ window

$I_{PMT}^{dark}$ - dark-current measured on the test PMT with the spectrometer lamp switched off

$I_{ref}^{'}(\lambda)$ and $I_{ref}(\lambda)$ - corresponding currents on the reference PMT measured with and without the $MgF_{2}$ window

$I_{ref}^{dark}$  - dark-current on the reference PMT with the spectrometer lamp switched off

\begin{figure}[t] % figures (and tables) should go top or bottom of
                    % the page where they are first cited or in
                    % subsequent pages
\begin{center}%
\subfiguretopcaptrue
\subfigure[][] % caption for subfigure a
{
\label{subfig:cooling_scheme}
\includegraphics[width=5cm]{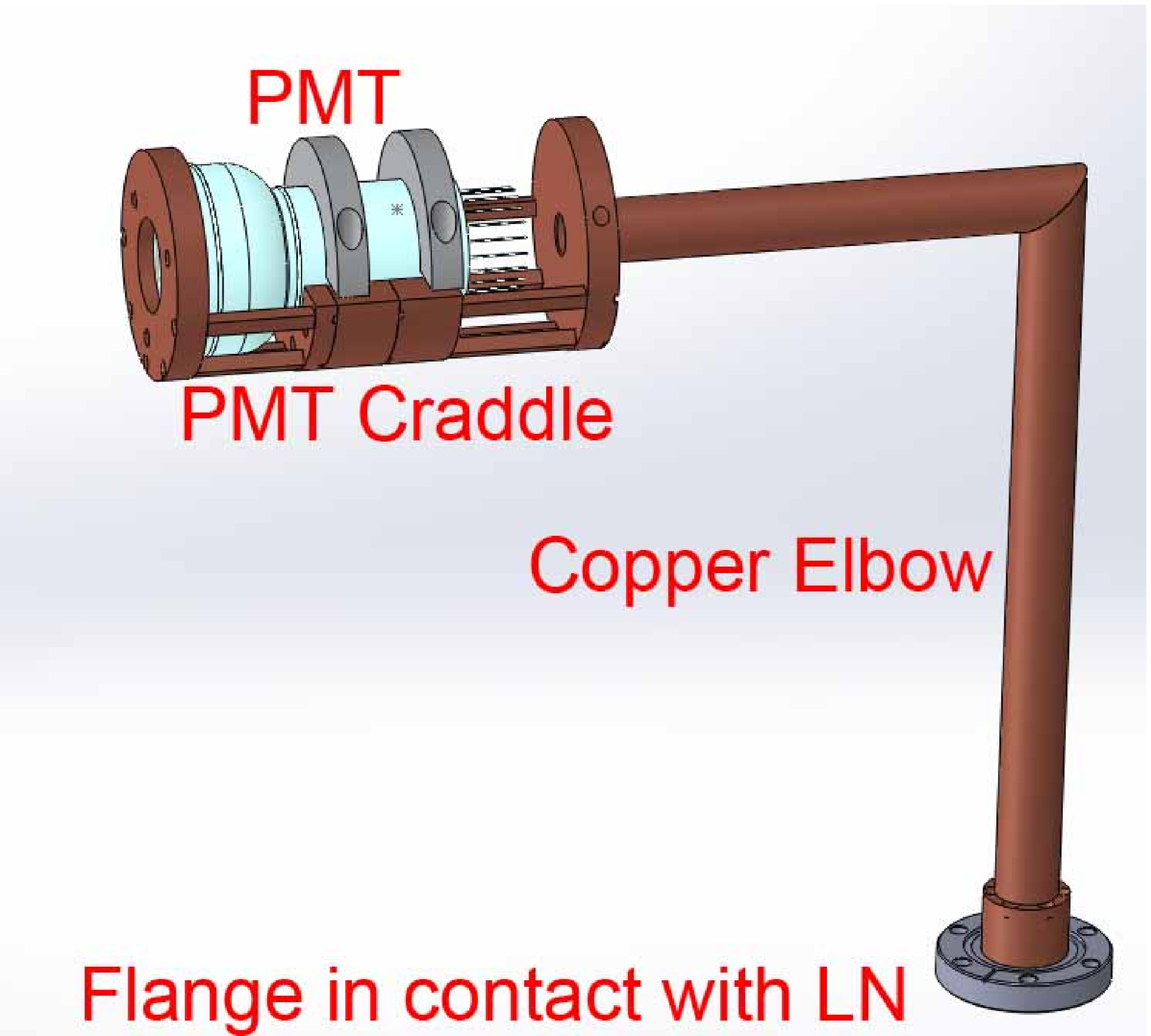}
}
\subfigure[][]
{
\label{subfig:cradle}
\includegraphics[height=5cm]{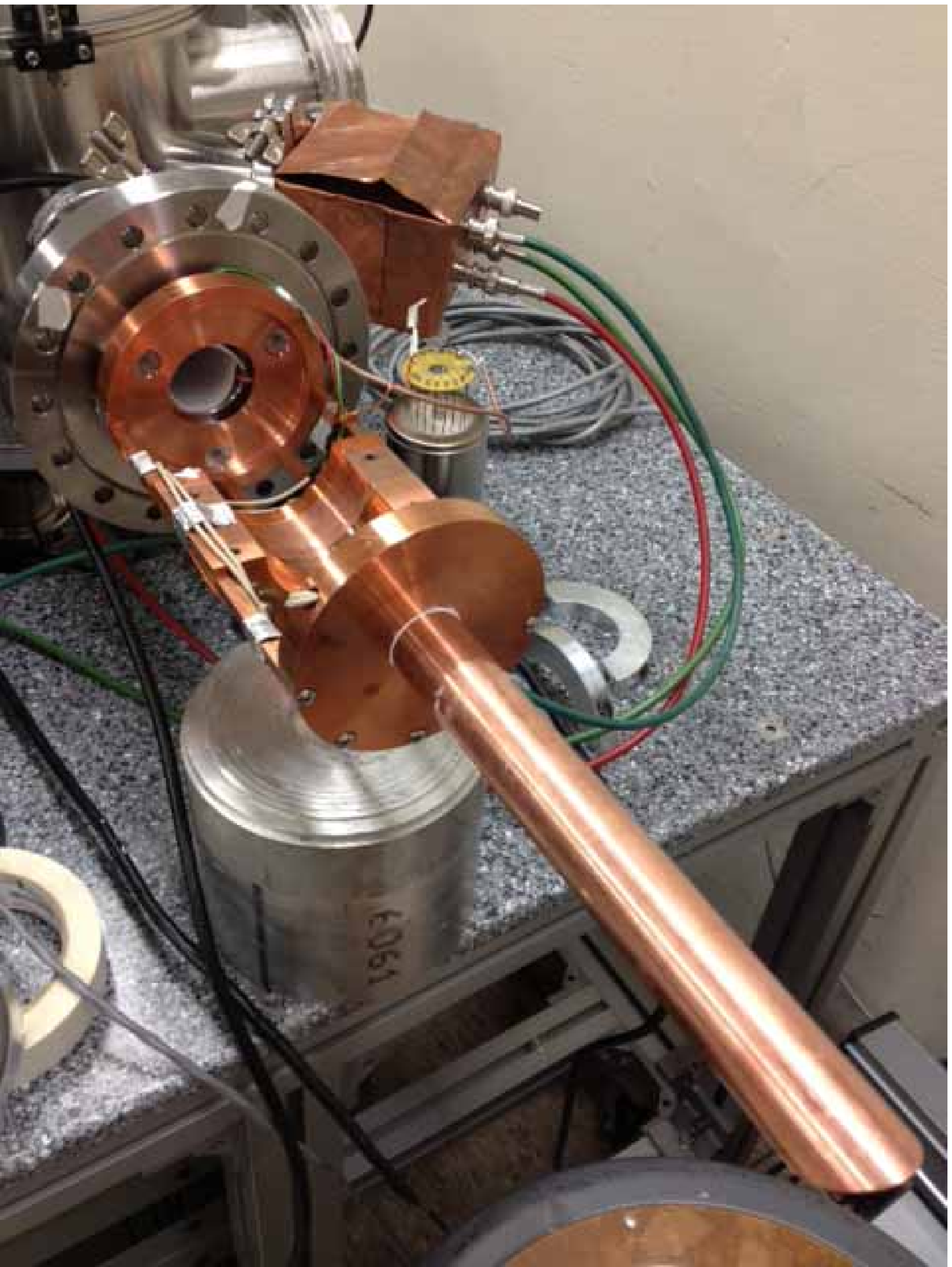}
}
\subfigure[][]
{
\label{subfig:pmt_in_cradle}
\includegraphics[width=5cm]{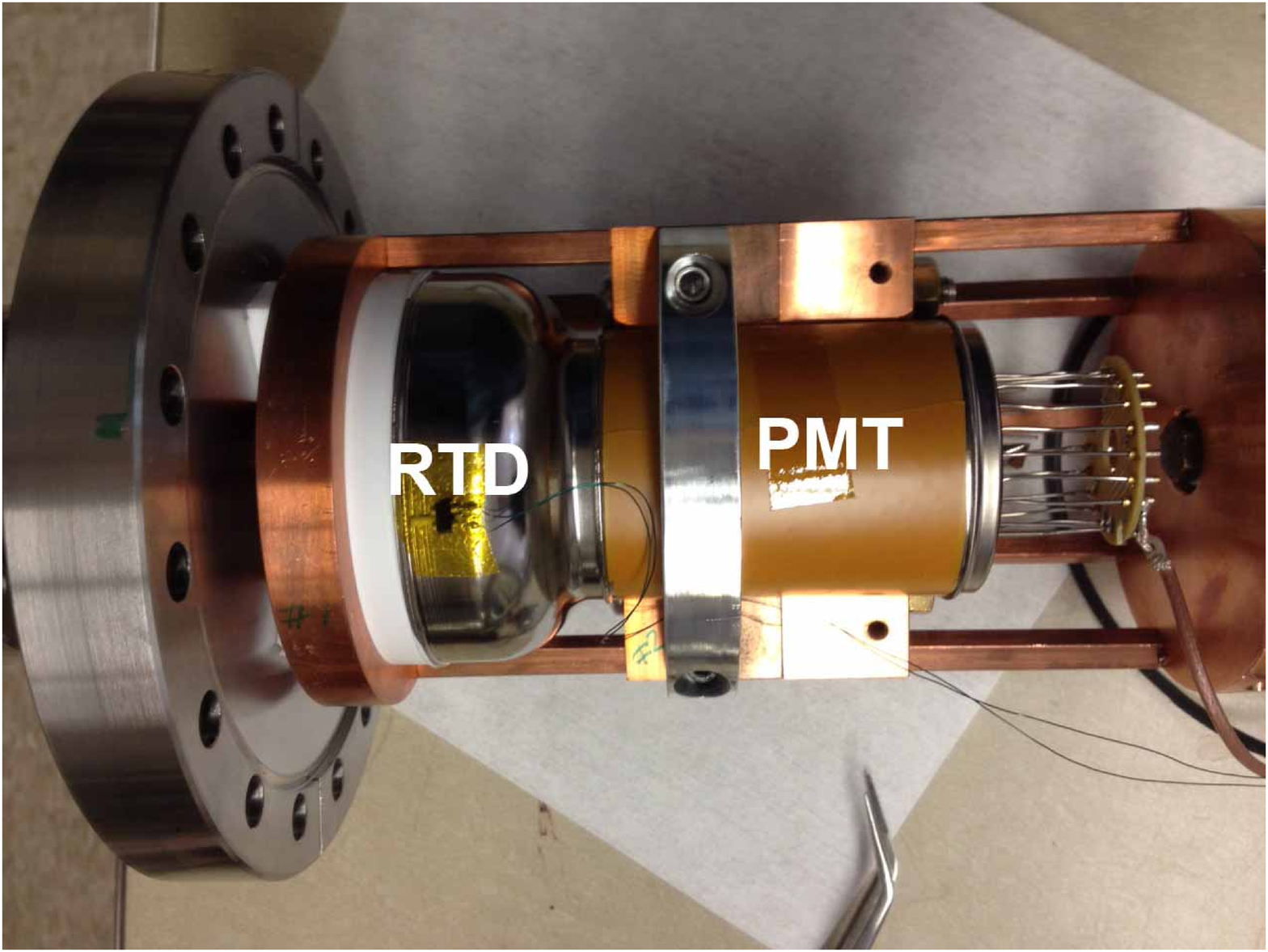}
}
\subfigure[][]
{
\label{subfig:cooling_2}
\includegraphics[height=7cm]{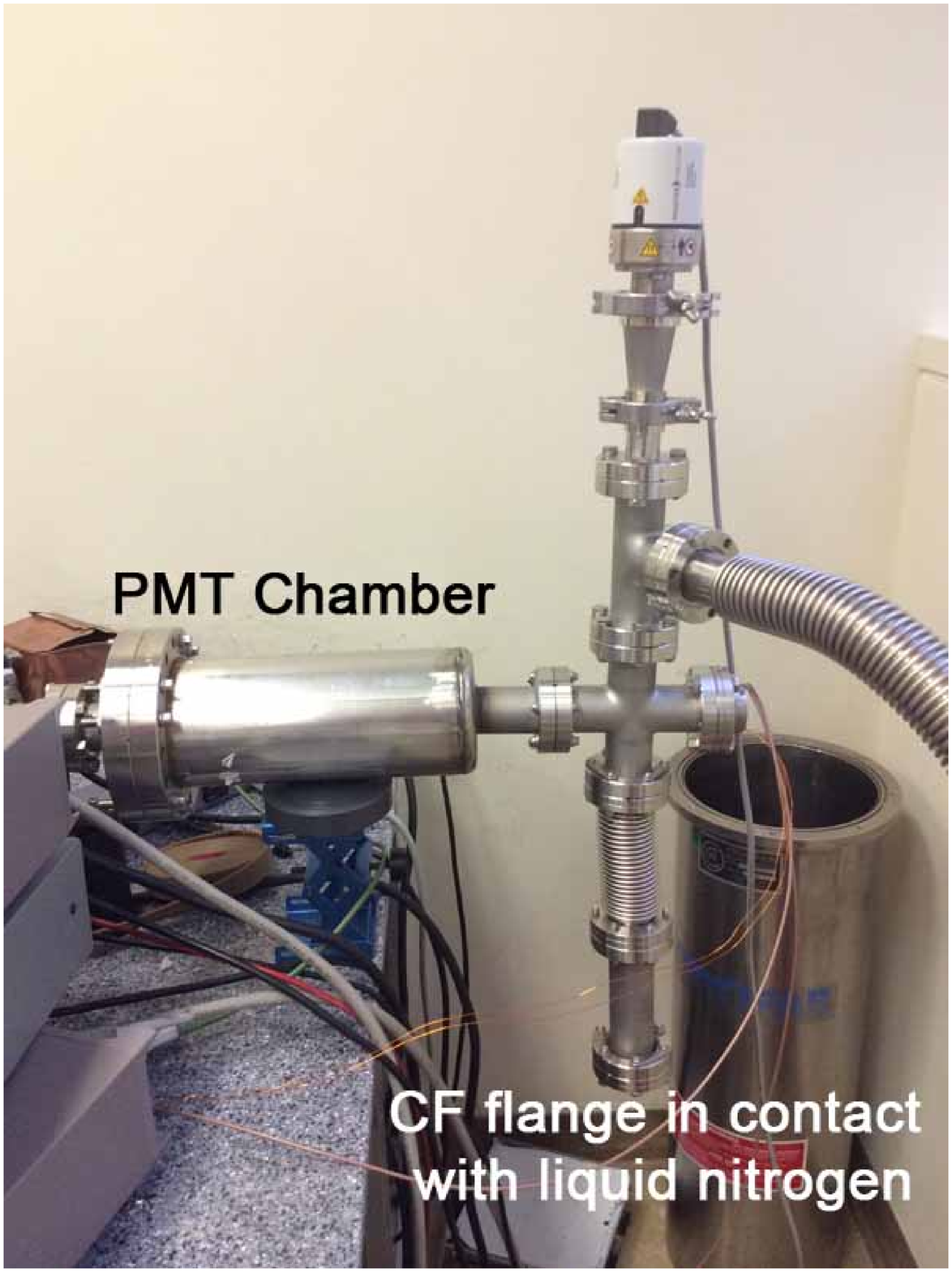}
}
\subfigure[][]
{
\label{subfig:cooling_1}
\includegraphics[height=7cm]{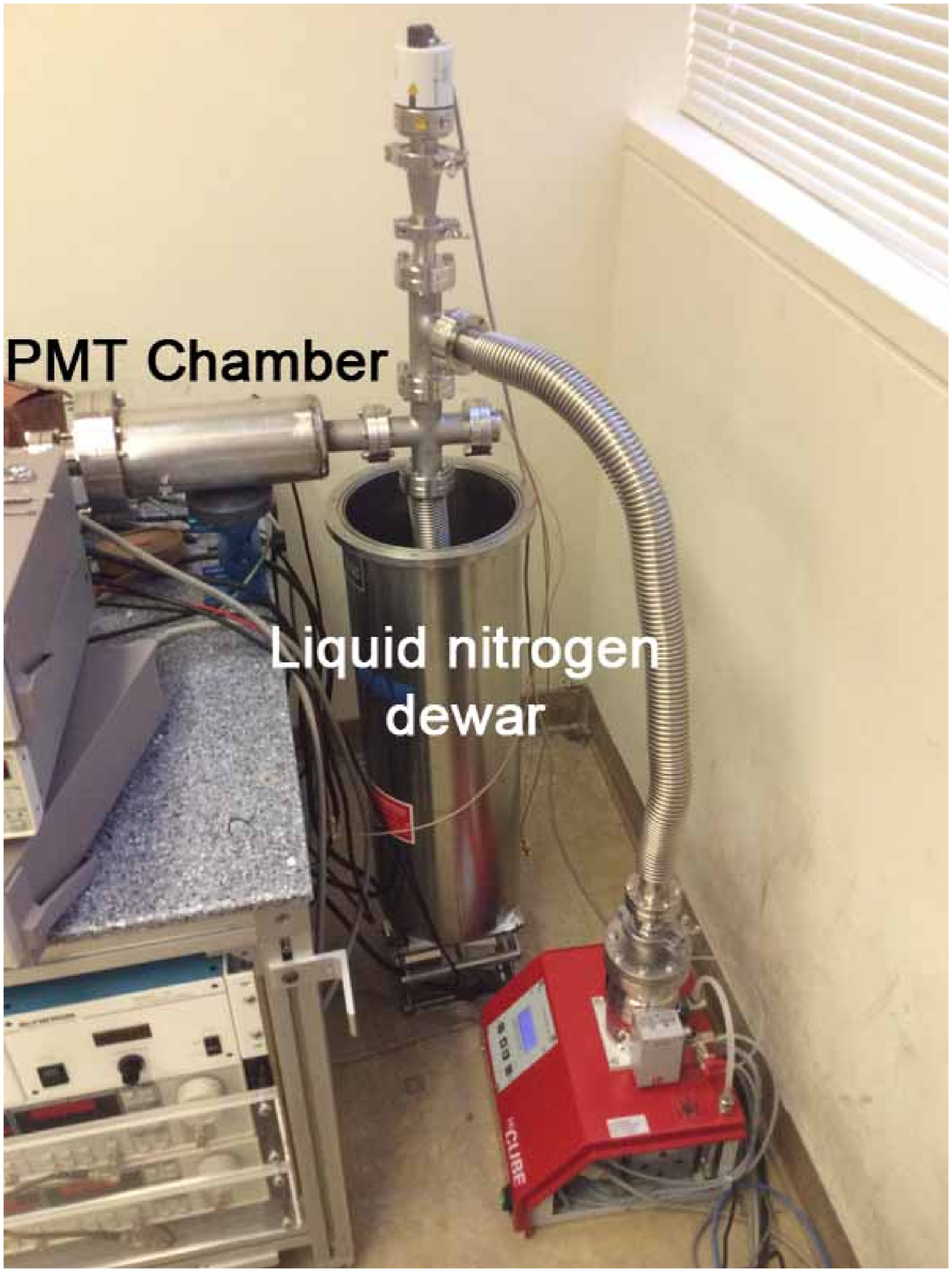}
}
\caption{PMT cooling system. a) Drawing of the PMT cooling system assembly without vacuum enclosure, b) Copper cradle for PMT cooldown, c) Hamamatsu R11410-10 PMT sitting in the cradle. One can see the RTD temperature sensor taped to the PMT case and Kapton PMT wrapping for electrical insulation; d) General view of the PMT cooling assembly inside a vacuum jacket; same as e) in a liquid Nitrogen Dewar for cooing.}
\label{fig:pmt_cooling}
\end{center}%
\end{figure}

Detailed photographs of the cooling system are presented in \ref{fig:pmt_cooling}. For the cooldown the test PMT was inserted into a copper cradle that was in thermal contact with a liquid nitrogen bath with an elbow made of 25.4 mm in diameter copper rod as shown in \ref{fig:pmt_cooling}. One end of the elbow was attached to the cradle while the other end was fixed at a $2-\frac{3}{4}$ inches conflate flange which was in direct contact with liquid Nitrogen (see \ref{subfig:cooling_scheme} and \ref{subfig:cradle}).

The test PMT was electrically isolated from the copper cradle with 50$\mu$m thick Kapton sheet (\ref{subfig:pmt_in_cradle}). The temperature on the test PMT was read out with a Resistive Temperature Detector (RTD) type Pt100 taped to the test PMT envelope next to the entrance window using a Kapton adhesive tape as shown in \ref{subfig:pmt_in_cradle}. That minimizes a thermal lag as the envelope is in direct thermal (and electric) contact with the entrance window of the test PMT. The PMT cradle was bolted to the front flange of PMT Chamber with plastic screws. It was thermally isolated from the flange using PTFE spacers as shown in \ref{subfig:pmt_in_cradle}. The copper assembly was then enclosed in a vacuum jacket for thermal insulation (see \ref{subfig:cooling_2} and \ref{subfig:cooling_1}).

The main sources of uncertainty in the measurements were the calibration uncertainty of the PD and the fluctuations in the PMT dark currents.

\section{Results}\label{sec:results}

\begin{figure}[tbp] % figures (and tables) should go top or bottom of
                    % the page where they are first cited or in
                    % subsequent pages
\begin{center}%
\includegraphics[width=9cm]{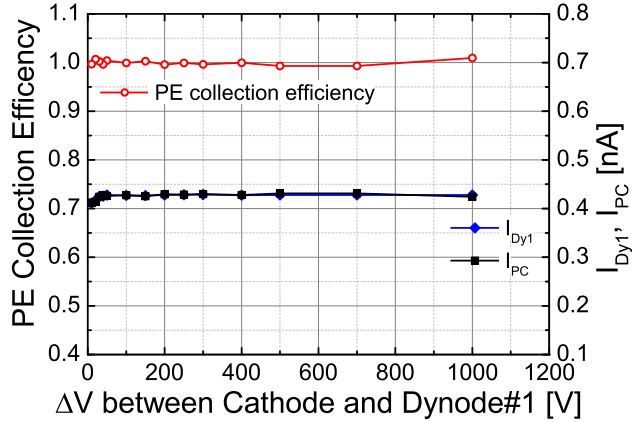}
\caption{Photoelectron collection efficiency as a function of voltage difference between the first dynode and the cathode of the test PMT. The currents measured on the cathode and on the anode of the test PMT are also shown.}
\label{fig:pe_coll_eff}
\end{center}%
\end{figure}

Before proceeding with the results of the absolute QE measurements we will demonstrate that the test PMT reaches nominal photoelectron collection efficiency at a positive voltage of 300V applied to its first dynode (which is interconnected with the PMT focusing grid). The photoelectron collection efficiency as a function of voltage difference between the cathode and the first dynode is shown in \ref{fig:pe_coll_eff}.

It can be seen from \ref{fig:pe_coll_eff} that the PMT shows full photoelectron collection efficiency already at a voltage difference of 10V between the first dynode and the cathode. The currents measured either from the first dynode or from the cathode reach a plateau after a voltage difference of about 30V between the first dynode and the cathode. Therefore, a voltage difference of 300V between the first dynode and the cathode ensures full photoelectron collection at the first dynode and the focusing grid.   

\begin{figure}[tbp] % figures (and tables) should go top or bottom of
                    % the page where they are first cited or in
                    % subsequent pages
\begin{center}%
\subfiguretopcaptrue
\subfigure[][] % caption for subfigure a
{
\label{subfig:qe_rt_ka28}
\includegraphics[width=9cm]{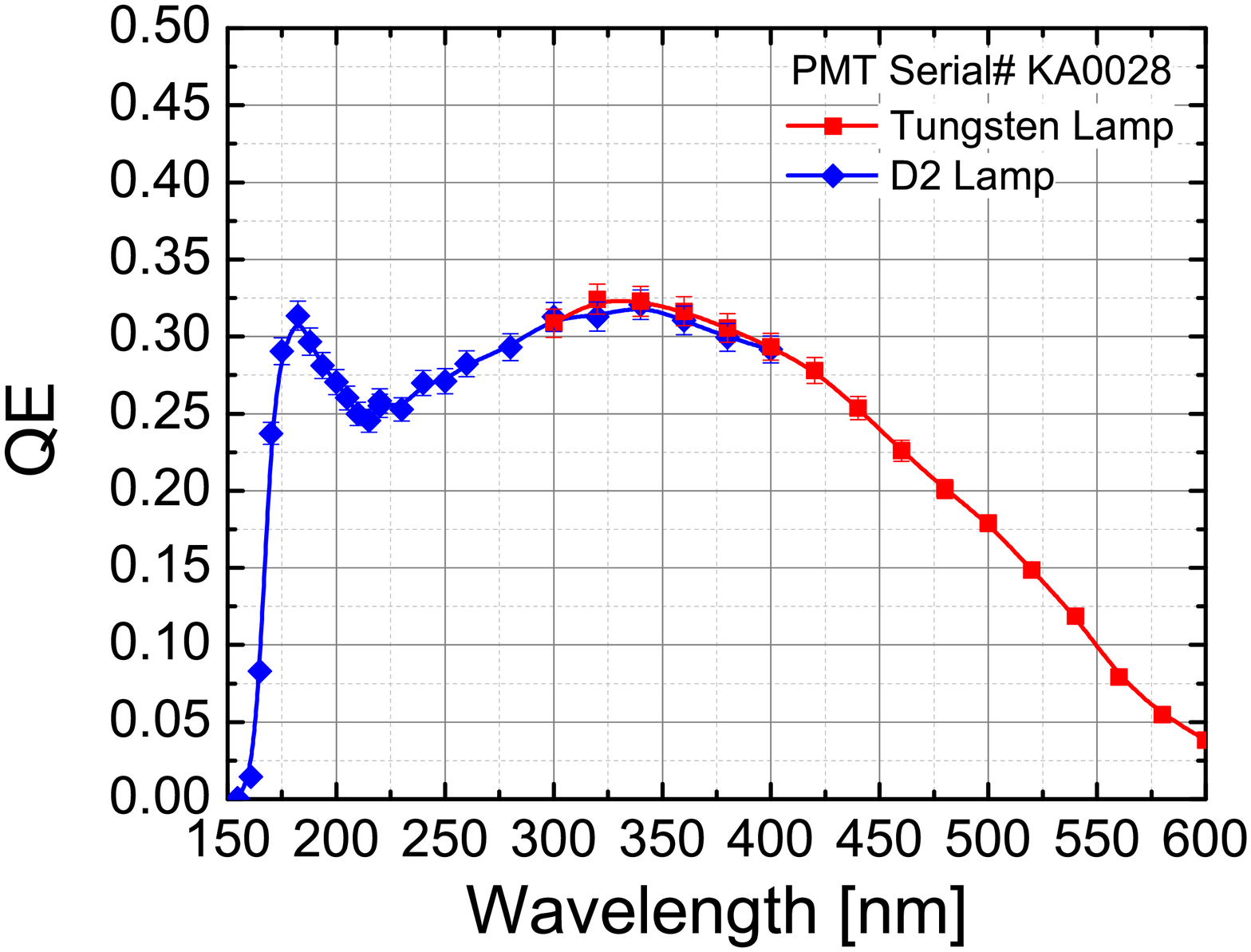}
}
\subfigure[][]
{
\label{subfig:qe_rt_ka44}
\includegraphics[width=9cm]{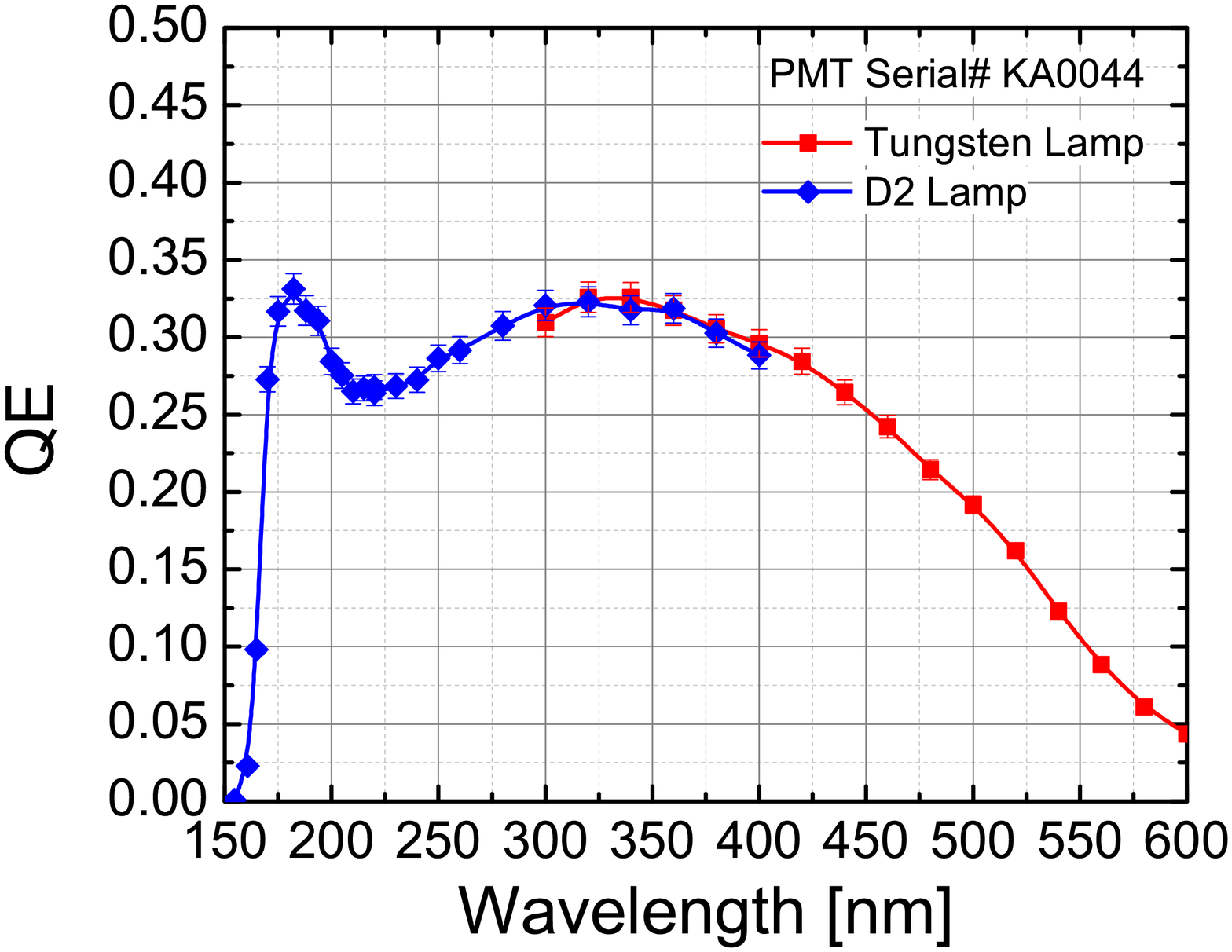}
}
\caption{Absolute QE measured at room temperature for Hamamatsu model R11410-10 PMTs serial \# a) KA0028 and b) KA0044. Red squares correspond to the measurements performed using a tungsten lamp as a light source, blue diamonds represent data points acquired using a deuterium lamp as a light source. }
\label{fig:qe_rt}
\end{center}%
\end{figure}

The absolute QE of the PMTs was measured first at room temperature. In \ref{fig:qe_rt} one can see two examples of such measurements; the QE is shown as a function of wavelength for PMTs serial \# KA0028 (\ref{subfig:qe_rt_ka28}) and KA0044 (\ref{subfig:qe_rt_ka44}). Blue diamonds represent the QE measured with the deuterium Lamp in a spectral range from 154.5 nm to 400 nm and red squares represent the QE measured with the tungsten lamp in a spectral range from 300 nm to 600 nm. In the plots we can see two characteristic peaks; first is at around 180 nm and the second one is at around 340 nm. Values of the absolute QE at 175 nm for PMT serial \# KA0028 and KA0044 were measured to be $29\%\pm0.9\%$ and $31.7\%\pm1\%$ correspondingly. Corresponding QE values provided by the manufacturer were 32.4\% and 33.9\% at 175 nm.

\begin{figure}[tbp] % figures (and tables) should go top or bottom of
                    % the page where they are first cited or in
                    % subsequent pages
\begin{center}%
\includegraphics[width=9cm]{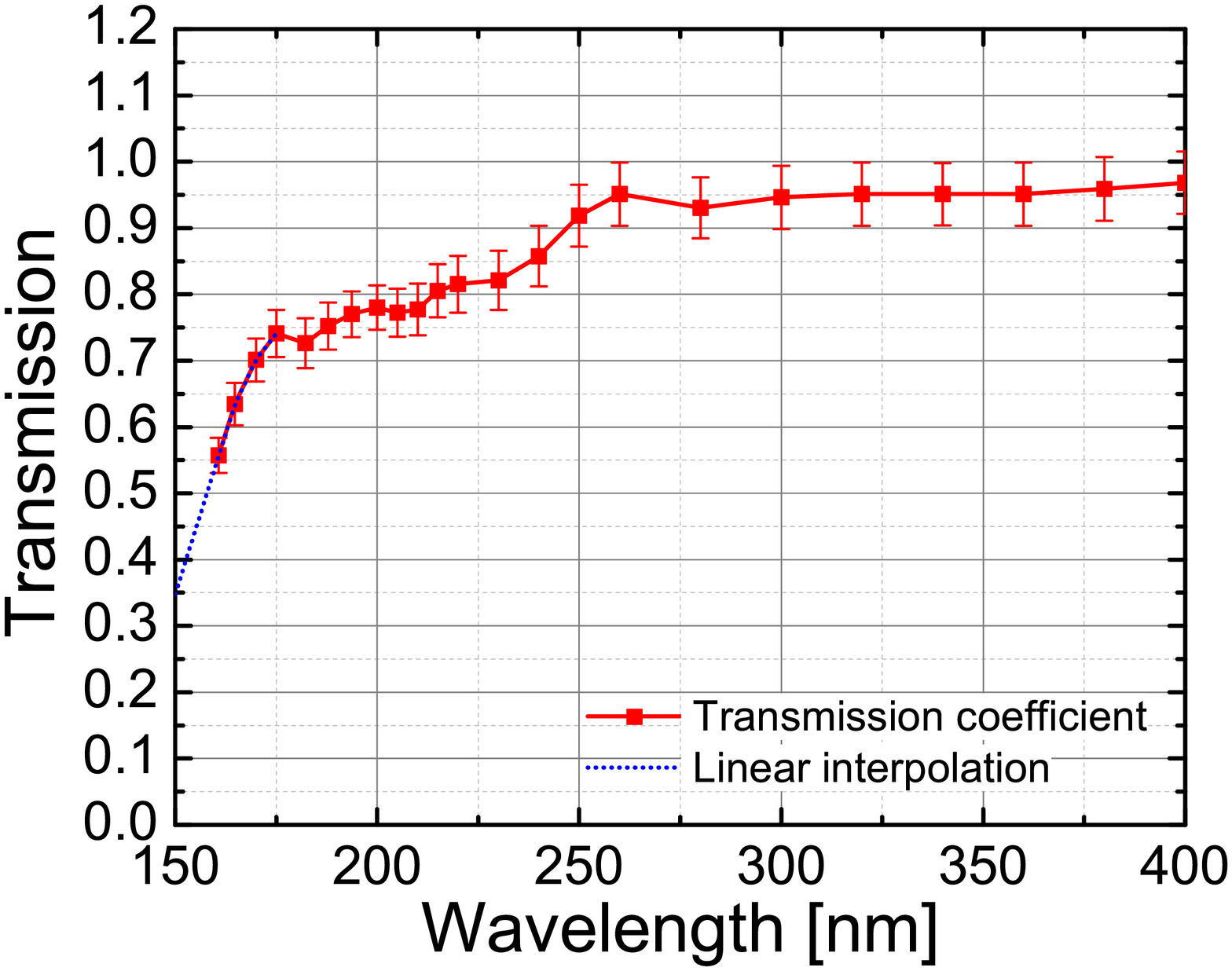}
\caption{Transmission coefficient of the $MgF_{2}$ window used in the QE measurements. Values of transmission coefficient below were estimated by linear extrapolation (blue dotted line) of data between 160.8 nm and 175 nm}
\label{fig:mgf2}
\end{center}%
\end{figure}

As mentioned in Section \ref{sec:exper}, the PMT chamber was separated from the PD Chamber with the $MgF_{2}$ window. The transmission coefficient for the $MgF_{2}$ window calculated using Eq. \ref{eq:trans} is presented in \ref{fig:mgf2} as a function of wavelength. It is required for calculation of the absolute QE the test PMT (see Eq. \ref{eq:qe}). At 154.5 nm the transmission coefficient was estimated using linear extrapolation of data between 160.8 nm and 175 nm due to a large measurement uncertainty caused by very low (comparable to dark current) signal current at the PMT. As we can see in \ref{fig:mgf2}, the transmission coefficient for the $MgF_{2}$ window below 250 nm is somewhat lower than that of commercially available VUV grade $MgF_{2}$ windows. This can be attributed to the surface contamination of the window after an extensive use.  

\begin{figure}[tbp] % figures (and tables) should go top or bottom of
                    % the page where they are first cited or in
                    % subsequent pages
\begin{center}%
\includegraphics[width=9cm]{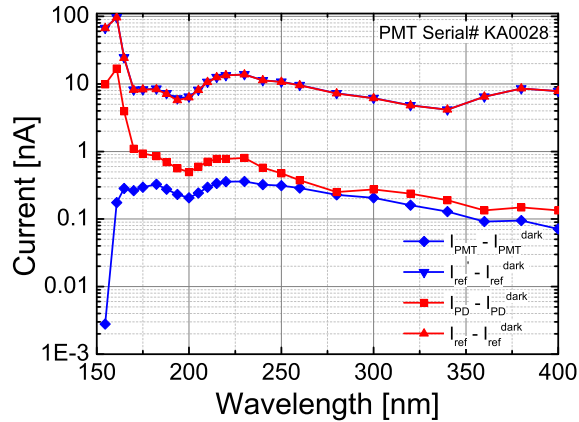}
\caption{An example of measured photocurrents as a function of wavelength. The annotations used in the legend correspond to those used in the Eq. \protect \ref{eq:qe}.}
\label{fig:photocurrents}
\end{center}%
\end{figure}

Using the transmission coefficient presented in \ref{fig:mgf2}, the absolute QE at low temperature was measured as the PMT was cooling down. It was derived from the measured photocurrents using Eq. \ref{eq:qe}. An example of measured photocurrents as a function wavelength is shown in \ref{fig:photocurrents}. The annotations used in \ref{fig:photocurrents} correspond to those used in the Eq. \ref{eq:qe}. During the measurements shown in \ref{fig:photocurrents} the dark current of the test PMT was fluctuating from -4 pA to -8 pA; an average value of -6 pA was taken in the calculations. The dark current of the PD was stable at 36 pA. The dark current of the reference PMT was stable at 12 pA.

\begin{figure}[tbp] % figures (and tables) should go top or bottom of
                    % the page where they are first cited or in
                    % subsequent pages
\begin{center}%
\includegraphics[width=9cm]{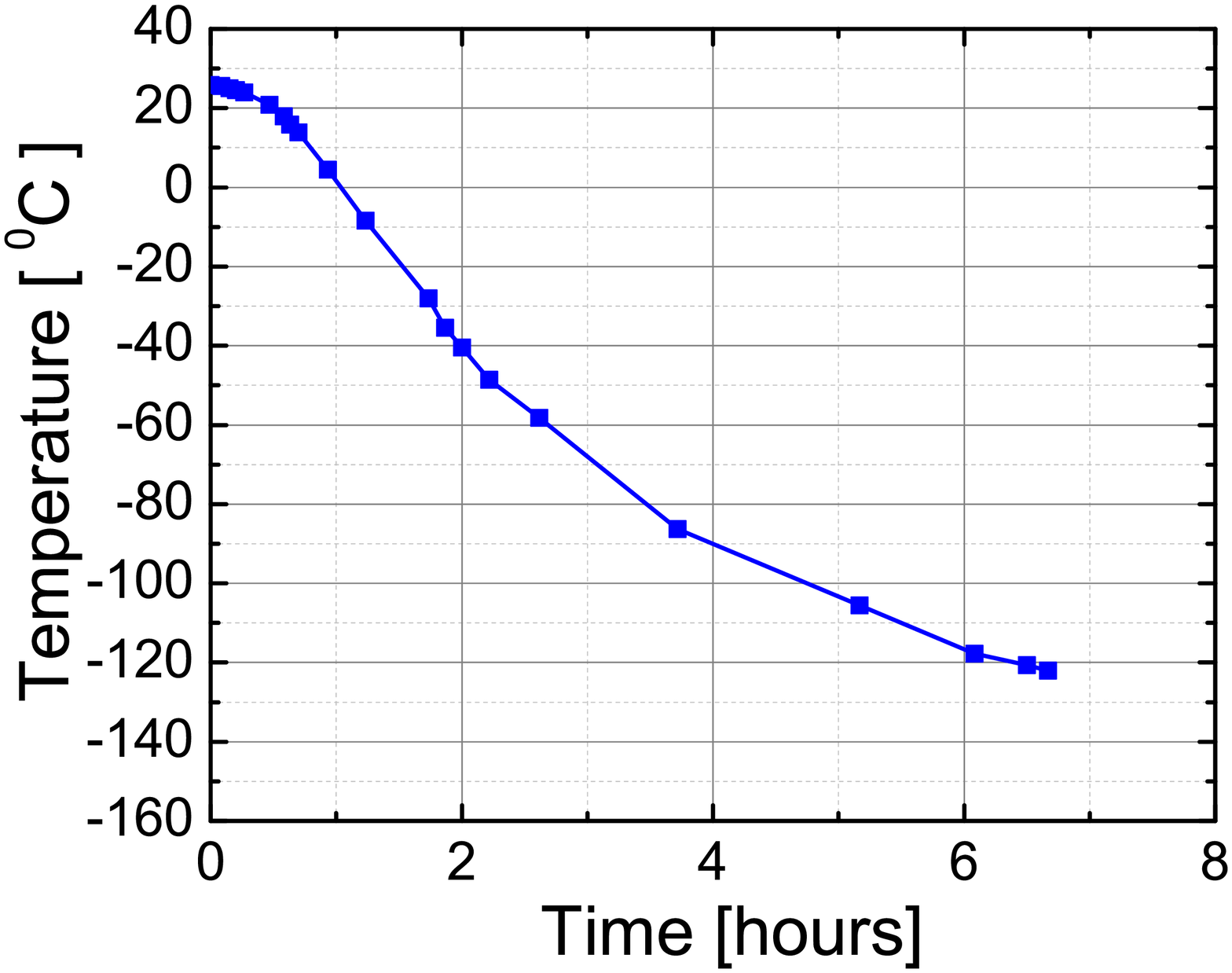}
\caption{Evolution of the PMT temperature with time during cooldown}
\label{fig:cooldown}
\end{center}%
\end{figure}

The evolution of the PMT temperature with time during the PMT cooldown from room temperature to $-120 ^{0}C$ is shown in \ref{fig:cooldown}. As seen in \ref{fig:cooldown} the PMT cools down to $-110 ^{0}C$ in approximately 5 hours which gives enough time for several QE measurements without any temperature control. As mentioned in Section \ref{sec:exper} during a QE measurement the temperature was varying by less than 2 $^{0}C$. Each PMT was subjected to one cooldown.

\begin{figure}[tbp] % figures (and tables) should go top or bottom of
                    % the page where they are first cited or in
                    % subsequent pages
\begin{center}%
\subfiguretopcaptrue
\subfigure[][] % caption for subfigure a
{
\label{subfig:qe_lt_ka07}
\includegraphics[width=9cm]{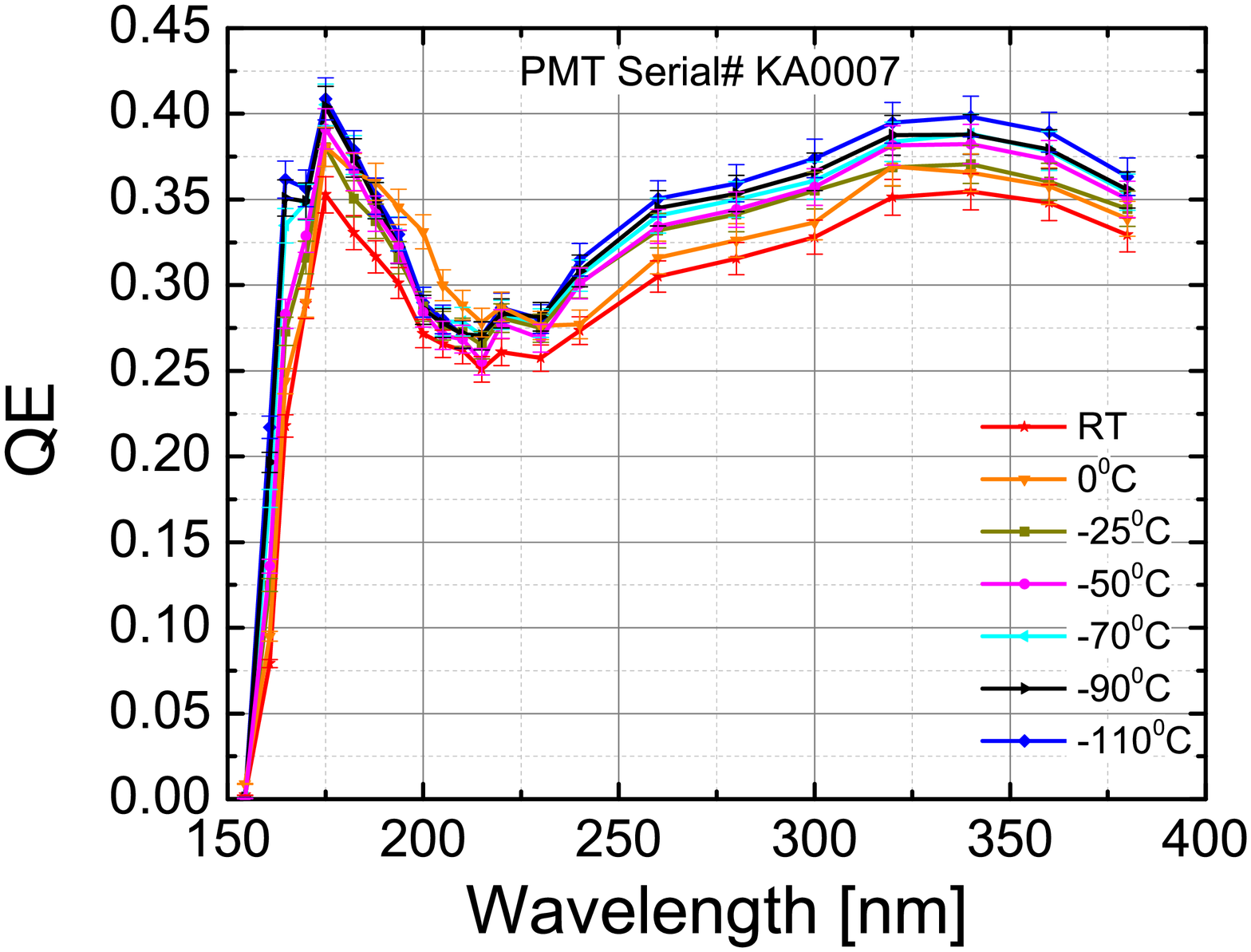}
}
\subfigure[][]
{
\label{subfig:qe_lt_ka35}
\includegraphics[width=9cm]{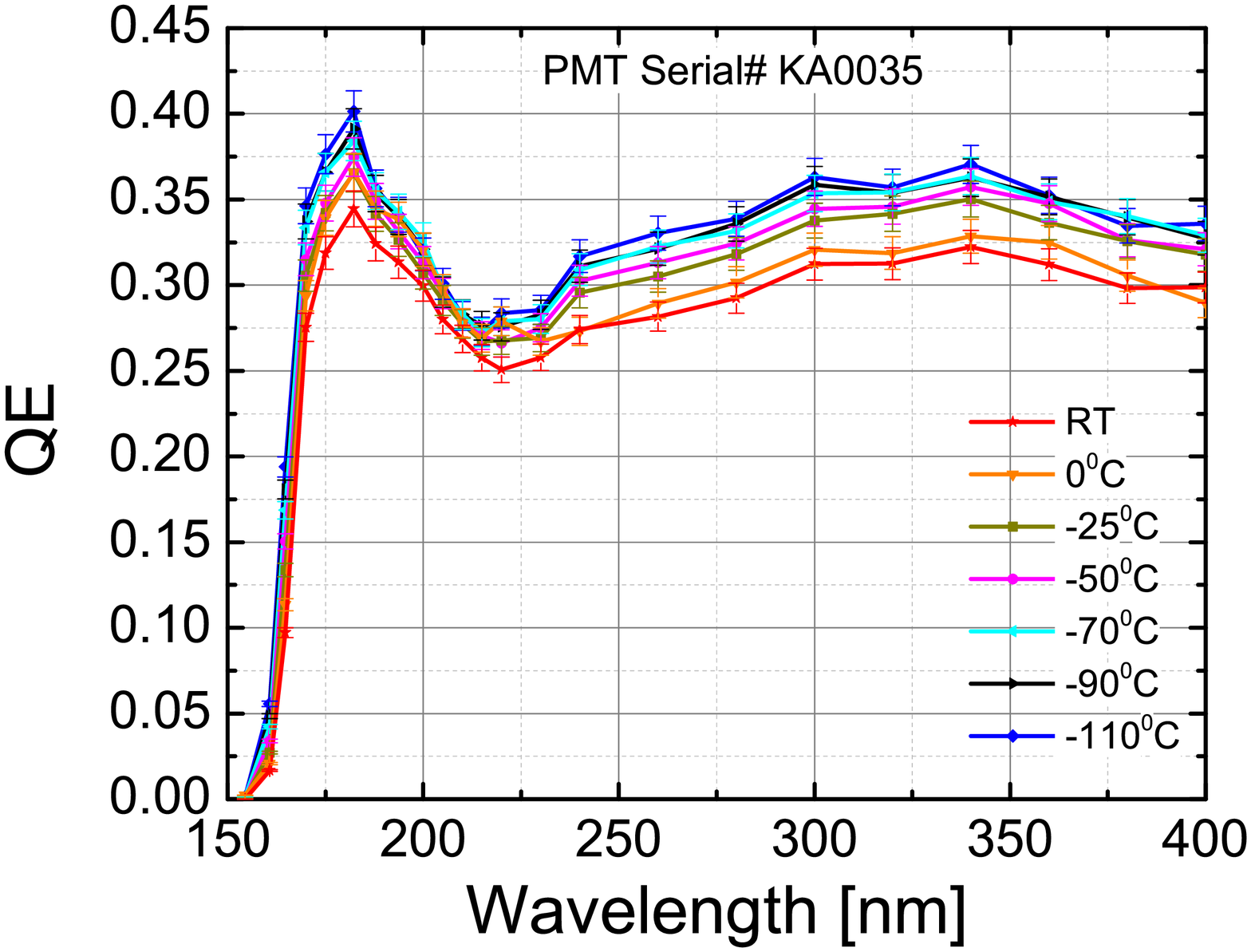}
}
\caption{Absolute QE as a function of wavelength measured at various temperatures for PMTs serial \# a) KA0007 and b) KA0035. It was recorded at the following temperatures: Room Temperature (red stars), 0 $^{0}C$ (orange triangles), -25 $^{0}C$ (dark yellow squares), -50 $^{0}C$ (magenta circles), -70 $^{0}C$ (cyan left arrows), -90 $^{0}C$ (black right arrows) and -110 $^{0}C$ (blue diamonds).}
\label{fig:qe_lowt}
\end{center}%
\end{figure}

In \ref{fig:qe_lowt} we present the absolute QE as a function of wavelength for Hamamatsu model R11410-10 PMTs measured at Room Temperature, 0 $^{0}C$, -25 $^{0}C$, -50 $^{0}C$, -70 $^{0}C$, -90 $^{0}C$ and -110 $^{0}C$ for PMTs serial\# KA0007 \ref{subfig:qe_lt_ka07},  KA0028 \ref{subfig:qe_lt_ka35}.

We learned from \ref{fig:qe_lowt} that during the cooldown from room temperature to -110 $^{0}C$ the absolute QE increases by about 10-15\% near the peaks. In the spectral range from 200 nm to 230 nm the QE variation with temperature is minor. The positions of the peaks slightly vary for different PMTs. In the case of PMT serial\# KA0007 (see \ref{subfig:qe_lt_ka07}) the first peak appears at around 175 nm, while for the rest of the PMT we tested the first peak is located at around 182 nm.

\begin{figure}[tbp] % figures (and tables) should go top or bottom of
                    % the page where they are first cited or in
                    % subsequent pages
\begin{center}%
\subfiguretopcaptrue
\subfigure[][] % caption for subfigure a
{
\label{subfig:qe_vs_t_175nm}
\includegraphics[width=9cm]{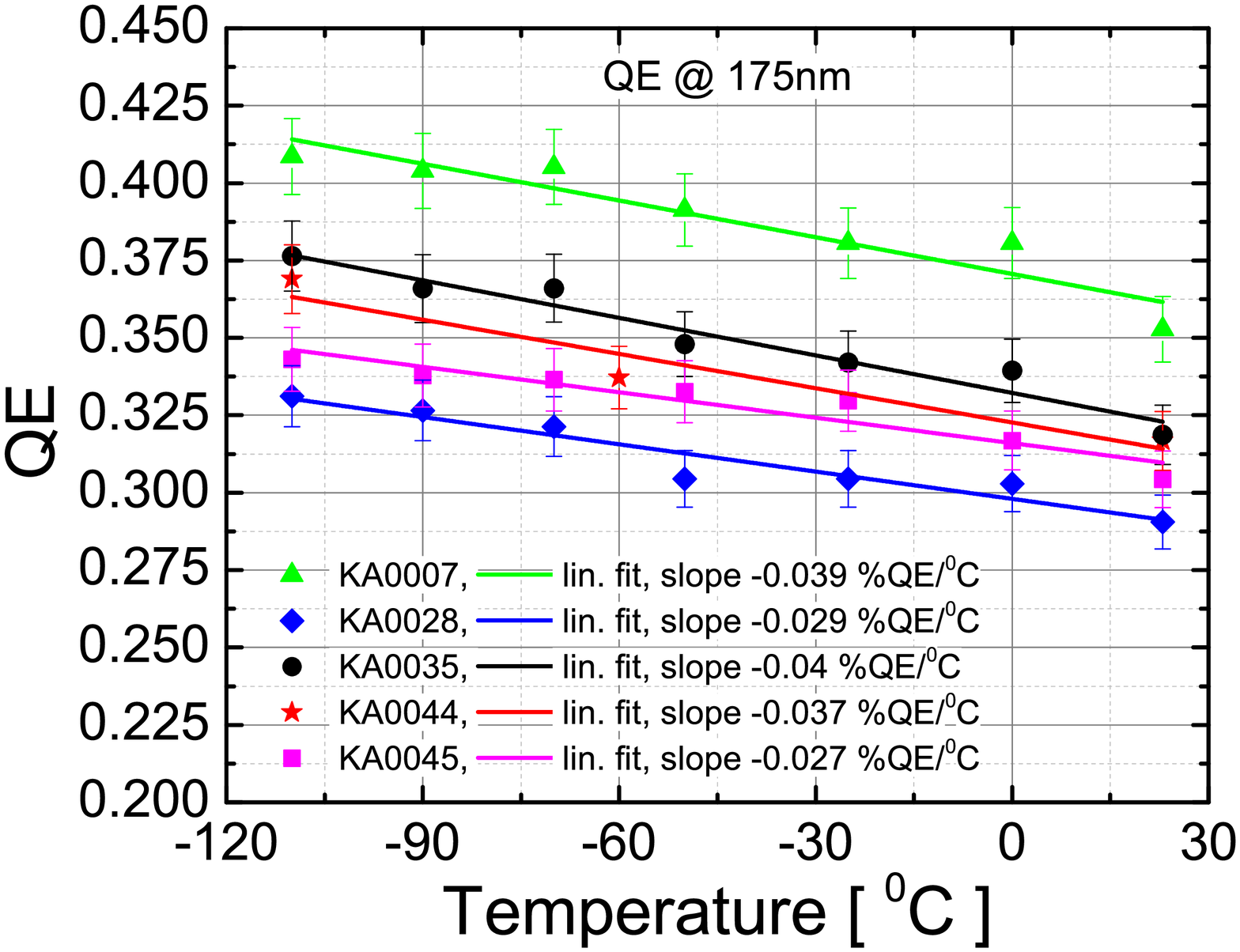}
}
\subfigure[][]
{
\label{subfig:qe_vs_t_340nm}
\includegraphics[width=9cm]{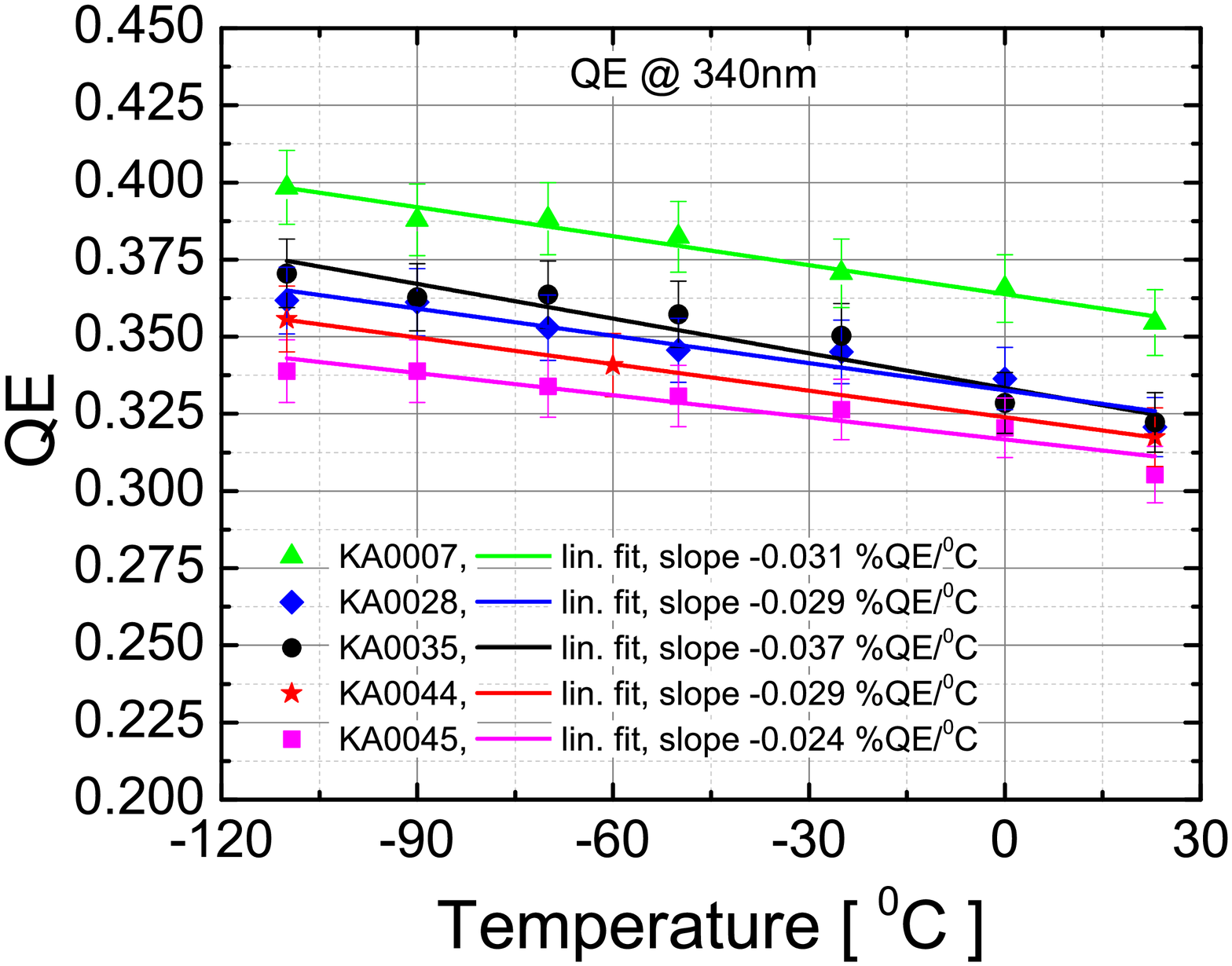}
}
\caption{Absolute QE as a function of temperature at a wavelength of a) 175 nm and b) 340 nm measured for the following PMTs serial\# KA0007 (green triangles), KA0028 (blue diamonds), KA0035(black circles), KA0044(red stars) and KA0045(magenta squares). Solid lines represent linear fits to each of the PMTs. The slope of the fitted line for each PMT is shown in the figure.}
\label{fig:qe_vs_t}
\end{center}%
\end{figure}

In \ref{fig:qe_vs_t} we showed how the absolute QE of the PMTs serial\# KA0007 (green triangles), KA0028 (blue diamonds), KA0035(black circles), KA0044(red stars) and KA0045(magenta squares) evolves with temperature at 175 nm (\ref{subfig:qe_vs_t_175nm}) that corresponds to the mean xenon scintillation wavelength and at 340 nm (\ref{subfig:qe_vs_t_340nm}) that approximately corresponds to a position of the second peak seen in \ref{fig:qe_rt} and in \ref{fig:qe_lowt}. To calculate the QE derivative with respect to temperature data points for each PMT were fitted with a straight line. The negative slope of the line represents the QE derivative with respect to temperature or the QE increase per every $^{0}C$. It can be seen in \ref{fig:qe_vs_t} that slopes of the fitted lines vary for various PMTs. Slopes for the same PMT are wavelength dependant. The behaviour of the QE derivative with respect to temperature at various wavelength for various PMTs was studied by a linear fit similar to the one shown in \ref{fig:qe_vs_t} at every wavelength used in the QE measurements.

\begin{figure}[tbp] % figures (and tables) should go top or bottom of
                    % the page where they are first cited or in
                    % subsequent pages
\begin{center}%
\includegraphics[width=9cm]{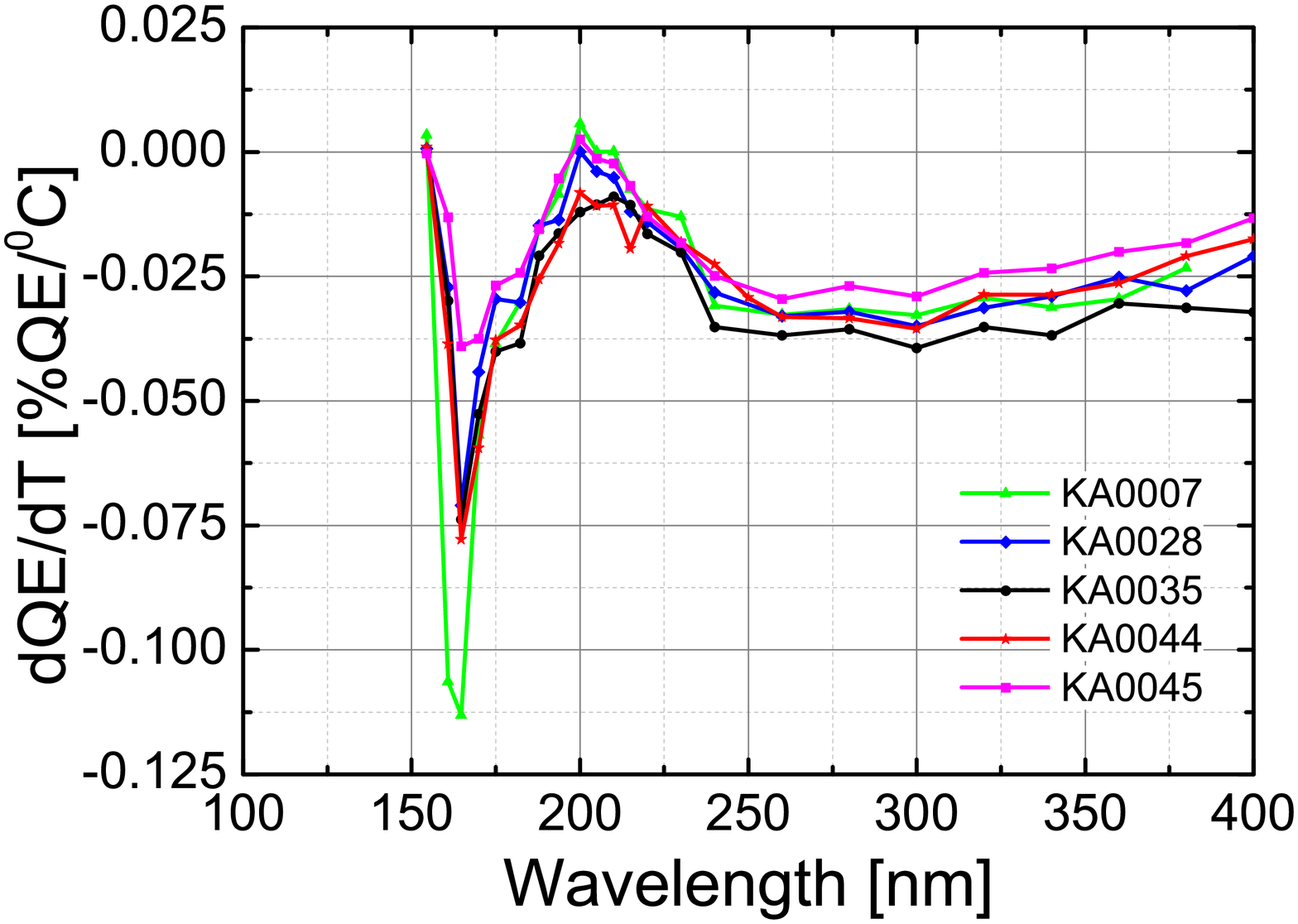}
\caption{QE growth rate with temperature as a function of wavelength or the QE derivative with respect to temperature measured for PMTs serial\# KA0007 (green triangles), KA0028 (blue diamonds), KA0035(black circles), KA0044(red stars) and KA0045(magenta squares)}
\label{fig:dqe_dt_vs_w}
\end{center}%
\end{figure}

The QE growth rate with temperature (measured in units \%QE/$^{0}C$) in this spectral range for all the PMTs used in the test is shown in \ref{fig:dqe_dt_vs_w} for the PMTs serial\# KA0007 (green triangles), KA0028 (blue diamonds), KA0035(black circles), KA0044(red stars) and KA0045(magenta squares). It can easily be seen in \ref{fig:dqe_dt_vs_w} that the QE variation rate with respect to temperature is not uniform over the spectral range as one can clearly see two characteristic peaks at around 165 nm and at around 200 nm. The first peak presents the fastest QE growth rate with decreasing temperature, while the second peak presents the slowest rate that in some cases approaches zero.  To further illustrate the meaning of \ref{fig:dqe_dt_vs_w} lets consider the QE growth rate with respect to temperature variation for PMT serial \# KA0035 (black circles). The first peak for this PMT appears around 165 nm having a value of -0.075\%QE/$^{0}C$. That means that if one decreases the temperature of the PMT serial \# KA0035 by 100 $^{0}C$ its QE will increase by 7.5\%. The value at the second peak at 210 nm for KA0035 is just about -0.01 \%QE/$^{0}C$ showing that the QE during cooldown to -100 $^{0}C$ will only gain about 1\% of its initial value.

\section{Conclusions and Discussion}\label{sec:conclusions}

The absolute QE for Hamamatsu model R11410-10 PMTs designed for use in low background liquid xenon experiments, was measured at room temperature and at low temperature down to the liquid Xenon boiling point. It was found that the QE enhances at low temperature.

Room temperature measurements in a spectral range from 154.5 nm to 600 nm demonstrated that the curves of QE versus wavelength have similar characteristic shape as those provided by manufacturer \cite{bib1}. One can see two characteristic peaks, one at around 182 nm and the other one at around 340 nm. The positions of the peaks slightly varied between different PMT samples. The measured QE values at 175 nm were similar to those measured by the manufacturer within the measurement uncertainty.

The absolute QE of five Hamamatsu model R11410-10 PMTs was measured at low temperatures down to -110 $^{0}C$ in a spectral range from 154.5 nm to 400 nm. It was demonstrated that during the PMT cooldown from room temperature to -110 $^{0}C$ (operation temperature of the PMTs in liquid xenon detectors) the QE increases by a factor of 1.1-1.15 at 175 nm. The increase of the QE at low temperatures can be accounted for the reduced photo-electron energy losses in the bulk photocathode material due to decrease of optical phonon cross-section with the photo-electron \cite{araujo:03}.

It was shown that the QE variation with temperature is wavelength dependant. For all of the test PMTs, the fastest QE growth rate with respect to temperature was found at around 165 nm, while the slowest one was observed at around 200 nm. This could be attributed to the variation of the initial photo-electron energies inside the photocathode bulk.

In application to the liquid xenon dark matter detectors the increase of QE of the PMT at low temperature will result in the improved single photon sensitivity. This will allow for lowering of the energy threshold and therefore, improving the detector sensitivity to the low energy WIMPs.

\acknowledgments

The authors thank S. Pordes of Fermilab, for his useful suggestions and great support of the project. This work was supported by the U.S. Department of Energy (DOE) under the award number DE-SC-0009937.

\end{document}